\documentclass[twocolumn]{aastex631}  

\shorttitle{Global CME-driven oscillations in the outer solar corona }
\shortauthors{Silva et al.}

\begin{document}

\title{Global compressive oscillations in the outer solar corona driven by an extreme Coronal Mass Ejection}

\correspondingauthor{Suzana S. A. Silva}
\email{suzana.silva@sheffield.ac.uk}

\author[0000-0001-5414-0197]{Suzana S. A. Silva}
\affiliation{Plasma Dynamics Group, School of Electrical and Electronic Engineering, University of Sheffield, Sheffield, S1 3JD, UK
 \\}

\author[0000-0003-0150-9418]{J. J. Gonz\'{a}lez-Avil\'{e}s}
\affiliation{Escuela Nacional de Estudios Superiores (ENES), Unidad Morelia, Universidad Nacional Aut\'{o}noma de M\'{e}xico, 58190 Morelia, Michoac\'{a}n, M\'{e}xico
 \\}

\author[0000-0002-1859-456X]{Pete Riley}
\affiliation{Predictive Science Inc., 9990 Mesa Rim Rd., Ste. 170, San Diego, CA 92121, USA
 \\}

\author[0000-0002-9164-0008]{Michal Ben-Nun}
\affiliation{Predictive Science Inc., 9990 Mesa Rim Rd., Ste. 170, San Diego, CA 92121, USA
 \\}

\author[0000-0002-4971-5854]{Erico L. Rempel}
\affiliation{Department of Mathematics, Aeronautics Institute of Technology, Pra\c ca Marechal Eduardo Gomes 50, 12228-900, S\~ao Jos\'{e} dos Campos, SP, Brazil \\
}

\author[0000-0002-2938-180X]{Leonardo F. G. Batista}
\affiliation{Departamento de F\'isica, Universidade Federal do Cear\'a, 6030 Campus do Pici, 60455-900, Fortaleza, CE, Brazil \\
}

\author[0000-0002-9546-2368]{Gary Verth}
\affiliation{Plasma Dynamics Group, School of Mathematical and Physical Sciences, University of Sheffield, Sheffield S3 7RH, UK
 \\}
 
 \author[0000-0002-3066-7653]{Istvan Ballai}
\affiliation{Plasma Dynamics Group, School of Mathematical and Physical Sciences, University of Sheffield, Sheffield S3 7RH, UK
 \\}

\author[0000-0002-5230-0270]{Chia-Hsien Lin,}
\affiliation{Department of Space Science and Engineering, National Central University, Taoyuan, Taiwan}

\author[0000-0002-5082-1398]{Luiz A. C. A. Schiavo}
\affiliation{Department of Mathematics, Physics and Electrical Engineering, Northumbria University, Newcastle NE1 8ST, UK}

\author[0000-0002-0893-7346]{Viktor Fedun}
\affiliation{Plasma Dynamics Group, School of Electrical and Electronic Engineering, University of Sheffield, Sheffield, S1 3JD, UK
 \\}


\begin{abstract}
Coronal mass ejections (CMEs) are known drivers of large-scale waves in the low corona. However, wave dynamics in the extended corona and inner heliosphere remain largely unexplored. Here, we report the first observational and numerical evidence of coherent global compressive oscillations in the outer corona and inner heliosphere, revealed by white-light SOHO/LASCO C3 data and an MHD simulation. Analyzing the CME event of 2012 July 23 using Spectral Proper Orthogonal Decomposition (SPOD), we isolate two distinct wave signatures: (1) a directional fast-mode shock-like compressive wave that dissipates completely within ~3 hours, and (2) a large-scale global circular wavefront consistent with fast-mode MHD behavior, lasting ~7 hours and extending across the LASCO C3 field of view, marking the first detection of such a global oscillation. Our findings reveal a previously unrecognized component of CME-driven wave activity, providing new constraints on the dynamics of the extended corona and inner heliosphere.
\end{abstract}


\section{Introduction}

Coronal mass ejections (CMEs) are some of the most violent and energetic phenomena in the solar atmosphere, driving massive amounts of magnetized plasma into the heliosphere. Expanding into the corona, CMEs drive the solar wind through a magnetic field environment and are instrumental in the generation of various magnetohydrodynamic (MHD) wave modes. These waves provide valuable information on the dynamics of CMEs and are important in influencing space weather \citep{Chen2005,Chen2013}.

A key manifestation of CME activity in the low corona is the appearance of expanding, annular extreme-ultraviolet (EUV) waves, first detected by the Extreme Ultraviolet Imaging Telescope (EIT) instrument and now observed in unprecedented detail. These so-called ``EIT" or EUV waves typically propagate at speeds of 200-400~km~s$^{-1}$ but can reach or exceed 1000~km~s$^{-1}$ in extreme cases; see, e.g. \cite{Ballai2005,Thompson2009,Warmuth2011}. Although early studies strongly linked EUV waves to CMEs and only weakly to solar flares \citep{Biesecker_2002}, their physical nature remains debated. Some researchers interpret EUV waves as true fast-mode MHD waves, while others observe them as pseudo-waves, large-scale signatures of coronal restructuring driven by the expanding CME \citep{wills2007,wang2009}. Also, \cite{kwon2013} and \cite{Kwon_2014} support a hybrid view, in which EUV waves often embody both a fast-mode wavefront and a non-wave component associated with coronal magnetic reconfiguration.

Fast CMEs are also well known for generating shocks, typically observed as type II radio bursts, and propagating at velocities of 500--2000~km~s$^{-1}$, or higher in extreme events \citep{Gopalswamy2008}. The formation mechanisms for these shocks can vary, with piston-driven shocks occurring during the early impulsive CME phase and bow shocks forming ahead of the CME front as it propagates outward \citep{Vrsnak2008,Veronig_2010,Kwon_2014}. The interplay of CME kinematics, coronal magnetic topology, and wave dynamics determines the spatial structure and evolution of both the shock and the associated global disturbances.

Recent observations and simulations have further revealed that CME eruptions can excite propagating shocks and large-scale coherent coronal oscillations. 
For instance, \cite{Veronig_2010,Liu_2017,Gopalswamy_2016} demonstrate that CME-driven shocks and their global imprints can envelop the entire solar disk, sometimes persisting for hours, and cover the full solar corona, including polar coronal holes \citep{Liu_2018}. Such global responses can include compressive standing modes and oscillations, suggesting a more intricate dynamical coupling between CME-driven disturbances and the extended corona than previously appreciated.

In this Letter, we present the first combined observational and numerical evidence for the simultaneous excitation of a fast-mode compressive front consistent with a shock in the outer corona and a global circular wavefront in the outer corona and inner heliosphere, triggered by the extreme CME of 23 July 2012, a benchmark event for large-scale wave generation in the solar corona \citep{Russell_2013, Baker2013}. We isolated and characterized the dual-wave response to the CME using Spectral Proper Orthogonal Decomposition (SPOD) analysis of SOHO/LASCO C3 data and a 3D MHD simulation. Our findings highlight how the CME excites multiple MHD wave components, including localized compressive fronts and global coronal oscillations, as revealed through modal decomposition. This imposes new constraints on the structure and dynamical response of the large-scale corona and inner heliosphere.

\section{Data and Methods}

\subsection{The 2012 July 23 CME event}
\label{July_2012_CME}

The 2012 July 23 CME event originated from active region (AR) 11520 on the far side of the Sun near S17W141 (heliographic longitude $\sim 141^{\circ}$), producing a moderate solar energetic particle event. The eruption was directed away from Earth toward $125^{\circ}$ W and was observed from multiple viewpoints by SOHO and STEREO \citep{2013SpWea..11..585B}. STEREO-A detected the CME arrival approximately 19 hours later. \citet{2014EP&S...66..104G} estimated a CME's leading-edge speed of $\sim 2600$ km s$^{-1}$ using the Graduated Cylindrical Shell (GCS) model applied to combined SOHO and STEREO data. 


To suppress the noise and to strengthen the faint coronal structures in LASCO C3 SOHO coronal data, we first applied a median filter \citep{1163188} with a window size of $10 \times 10$ pixels to remove isolated noise and to keep physical structures such as high gradients or propagating wavefronts. Following this step, the temporal sequence of the images was smoothed in the time domain with a Savitzky-Golay filter \citep{Savitzky1964} (9 frame window) to filter out high-frequency temporal noise without sacrificing essential features. This preprocessing was necessary to perform a reliable study and avoid noise-related artifacts in the subsequent results.
\subsection{Spectral Proper Orthogonal Decomposition}

To identify coherent wave structures, we employed the SPOD method, which enhances the classical Proper Orthogonal Decomposition (POD) method by incorporating spectral filtering \citep{sieber2016spectral}. POD is a statistical method used to extract dominant spatial structures (modes) from time-dependent data by analyzing correlations in the temporal evolution of a physical field. Given a time-resolved field $q(\mathbf{x}, t)$, we define its fluctuation as
\begin{equation}
q^{\prime}(\mathbf{x}, t) = q(\mathbf{x}, t) - \left\langle q(\mathbf{x}) \right\rangle,
\end{equation}
where $\langle\cdot\rangle$ denotes the temporal mean.

For analysis of datacube images, the temporal covariance matrix is given by
\begin{equation}
C_{t_1,t_2} = \frac{1}{N} \int_{\Omega} q^{\prime}(\mathbf{x}, t_1) q^{\prime}(\mathbf{x}, t_2) \, d\mathbf{x},
\end{equation}
with $N$ being the number of snapshots and $\Omega$ the spatial domain,  SPOD applies a temporal convolution to $C$ using a Gaussian filter, $g_k$, yielding the spectrally filtered matrix
\begin{equation}
S_{i,j} = \sum_{k=-N_f}^{N_f} g_k \, C_{i+k, j+k},
\end{equation}
which improves mode separation in the frequency space \citep{Ribeiro_2017}. 

For the analysis of 3D MHD simulations, vectorial SPOD is more suitable, as it captures the joint dynamics of all vector field components. In this case, the temporal covariance matrix is defined as
\begin{equation}
C_{t_1,t_2} = \frac{1}{N} \int_{\Omega} {\bf v}^{\prime}(\mathbf{x}, t_1)\cdot{\bf v}^{\prime}(\mathbf{x}, t_2) \, d\mathbf{x},
\label{eq:vec-SPOD}
\end{equation}
where ${\bf v}$ is the velocity vector. 

In both cases, the decomposition of the singular value of $S$ results in orthonormal spatial modes, $\phi^{(n)}(\mathbf{x})$, reconstructed as
\begin{equation}
\phi^{(n)}(\mathbf{x}) = \frac{1}{\lambda_n N} \sum_{j=1}^{N} \xi_{n}(t_j) \, q^{\prime}(\mathbf{x}, t_j),
\end{equation}
where $\bm{\xi}_n$ and $\lambda_n$ are components of the eigenvector and eigenvalues of $\bm{C}\bm{\xi}_n = \lambda_n \bm{\xi}_n$.

The temporal amplitude of each mode is given by
\begin{equation}
\bm{a}^{(n)}(t) = \sqrt{N \lambda_n} \, \bm{\xi}_{n}.
\end{equation}
The temporal amplitude $\bm{a}^{(n)}(t) $ describes the time evolution of the $n$-th mode, enabling the identification of coherent wave dynamics and their temporal characteristics such as frequency, growth, or decay.

\subsection{Numerical Model}
\label{sec:models}

To simulate CME propagation through a steady-state solar wind, we used \texttt{sunRunner3D}, a three-dimensional MHD model that extends the earlier 1D version, \texttt{SunRunner1D} \citep{Riley&Ben-Nun_2022}. \texttt{SunRunner3D} has been earlier applied to model CME structures \citep{Riley_et_al_2025}, interpret the global heliospheric structure from in situ data using simulations \citep{Gonzalez-Aviles_et_al_2024}, assess numerical effects in solar wind simulations \citep{De_leon_alanis_et_al_2024}, and reproduce stream/corotating interaction regions observed by Parker Solar Probe (PSP) and STEREO-A \citep{Aguilar-Rodriguez_et_al_2024}.

Here, \texttt{sunRunner3D} employs boundary conditions from CORona-HEliosphere (CORHEL) \citep{2009AGUFMSA43A1612L} and solves the ideal MHD equations using the PLUTO code \citep{Mignone_et_al_2007}. The conservative form of the equations is 
\begin{eqnarray}
\frac{\partial\varrho}{\partial t} + \nabla\cdot(\varrho{\bf v}) &=& 0, \label{density}\\
\frac{\partial(\varrho{\bf v})}{\partial t} + \nabla\cdot(\varrho{\bf v}{\bf v}-{\bf B}{\bf B} + p_{t}{\bf I}) &=& \varrho{\bf g}, \label{momentum}\\
\frac{\partial E}{\partial t} + \nabla\cdot((E+p_{t}){\bf v} - {\bf B}({\bf v}\cdot{\bf B})) &=& \varrho{\bf v}\cdot{\bf g}, \label{energy}\\
\frac{\partial{\bf B}}{\partial t} + \nabla\cdot({\bf v}{\bf B} - {\bf B}{\bf v}) &=& 0, \label{evolB}\\
\nabla\cdot{\bf B} &=& 0, \label{divB}
\end{eqnarray} 
where $\varrho$ is the mass density, ${\bf B}$ magnetic field, and $p_{t} = p + B^{2}/2$ the total pressure, and ${\bf I}$ is the identity matrix. The total energy is $ E = p/(\gamma-1) + \varrho v^{2}/2 + B^{2}/2 $, with $\gamma = 5/3$,  $p = \varrho k_B T/\bar{m}$. Here, $T$ is the temperature of the plasma, $\bar{m}=\mu m_{p}$ is the particle mass specified by a mean molecular weight value $\mu = 0.6$ for a fully ionized gas, $m_{p}$ is the proton mass, and $k_B$ is the Boltzmann constant. Gravity is modeled as ${\bf g}=-GM_\odot/r^2 \, \hat{\bf r}$, with $G$ representing the gravitational constant and $M_{\odot}$ the solar mass. In this paper, we show the results in terms of the proton number density $N_p$ and proton temperature $T_p$, instead of the mass density $\rho$ and plasma temperature $T$, as appropriate for a fully ionized hydrogen plasma in which protons dominate both mass and thermal properties.

Equations (\ref{density})-(\ref{divB}) are solved in 3D spherical coordinates $(r,\theta,\phi)$, including solar rotation in longitude ($\phi$) at a rate equal to $\Omega_c = 2.8653 \times 10^{-6}$ Hz \citep{Pomoell&Poedts_2018, Mayank_2022}. The domain spans 0.14–1.1 AU in $r$, $\theta\in[0,\pi]$ and $\phi\in[0,2\pi]$, with resolution $141\times111\times128$. We use RK2 time-stepping, second-order reconstruction with minmod limiter, the HLL Riemann solver \citep{Harten1997}, and Powell’s eight-wave method \citep{Powell1997} for magnetic divergence control. Boundary conditions include CORHEL input  at 0.14 AU, outflow at 1.1 AU, polar axis treatment with the ring-average method in $\theta$ \citep{ZHANG2019276}, and periodicity in $\phi$.

\subsubsection{CME model}
\label{CME_model}

We modeled the CME using a cone model, a typical approximation commonly used to describe the properties of CMEs in the heliosphere \citep{Odstrcil_et_al_2004, Pomoell&Poedts_2018}. In our model, the CME is initialized as a circular perturbation with time-dependent angular width and radius that evolve sinusoidally. Accordingly, we use

\begin{eqnarray}
& &\left(\phi-\phi_{\text{CME}}\right)^2 + \left(\theta-\theta_{\text{CME}}\right)^2 < \alpha(t)^2, \\
&&\alpha(t) = \omega_{\text{CME}} \sin\left[\frac{\pi}{2} \frac{(t - t_{\text{onset}})}{t_{\text{half}}} \right], \\
&&t_{\text{half}} = R_{\text{in}} \frac{\tan(\omega_{\text{CME}}/2)}{v_{\text{CME}}},
\end{eqnarray}
where $\phi_{\text{CME}}$ and $\theta_{\text{CME}}$ are the CME center coordinates, $\alpha(t)$ the angular extent, $\omega_{\text{CME}}$ the angular width, $t_{\text{onset}}$ the onset time, $t_{\text{half}}$ the insertion half-time, $ R_{\text{in}} = 0.14$ AU, the location of the inner boundary, and $v_{\text{CME}}$ the CME speed.

To simulate the 2012 July 23 CME event in \texttt{sunRunner3D}, we extracted event parameters from the DONKI database\footnote{\url{https://kauai.ccmc.gsfc.nasa.gov/DONKI/}} and LASCO coronagraph data. CME speed at 31 $R_{\odot}$ was obtained from a second-order fit in the SOHO LASCO CME Catalog\footnote{\url{https://cdaw.gsfc.nasa.gov/CME_list/}}. The values of $\phi_{\text{CME}}$ and $\theta_{\text{CME}}$ in the Heliocentric Earth Equatorial (HEEQ) coordinates were converted to the Carrington coordinates using \texttt{SunPy} to match our simulation framework.

In summary, the input parameters used for this event are: detection time = 2012-07-23T02: 36:00, $ t_{\text{onset}} = 240$ h, $t_{\text{half}} = 120$ h, $\theta_{\text{CME}} = 75.0^{\circ}$, $ \phi_{\text{CME}} = 87.2^{\circ} $, $ \omega_{\text{CME}} = 88.0^{\circ} $ and $ v_{\text{CME}} = 2000$ km s$^{-1}$.

\subsection{Synthetic White light}
To compute simulated brightness (B) and polarized brightness (pB) images from the simulation results, we use a forward modeling technique based on the Thomson scattering formulation originally described by \cite{1966gtsc.book.....B}. In particular, \texttt{sunRunner3D} provides full 3-D electron density distributions, which we integrate along lines of sight to synthesize the observed white-light emission. The total brightness is calculated by summing the scattered intensity from all electrons along the line of sight, taking into account the geometry-dependent scattering efficiency. Polarized brightness is derived by computing the component of scattered light polarized perpendicular to the scattering plane, which emphasizes density structures nearer the plane of the sky. We implement these calculations using a tool, getpb, which projects the model results into the observer's geometry, including spacecraft or ground-based eclipse locations, and applies the appropriate scattering physics. This approach allows us to generate synthetic B and pB data for direct comparison with coronagraph, heliospheric images, or eclipse observations. These kind of data closely resemble those expected from the upcoming \textit{Polarimeter to UNify the Corona and Heliosphere (PUNCH)} mission \citep{2022tess.conf12522K}.

The method follows the foundations developed by \cite{1966gtsc.book.....B} and includes refinements from more recent treatments, such as those by \cite{Howard_2012}, who reviewed practical implementation considerations for heliospheric imaging and extended Thomson scattering theory to modern instrumentation and viewing conditions.

\section{Results}

\begin{figure}
    \centering
    \includegraphics[width=\linewidth]{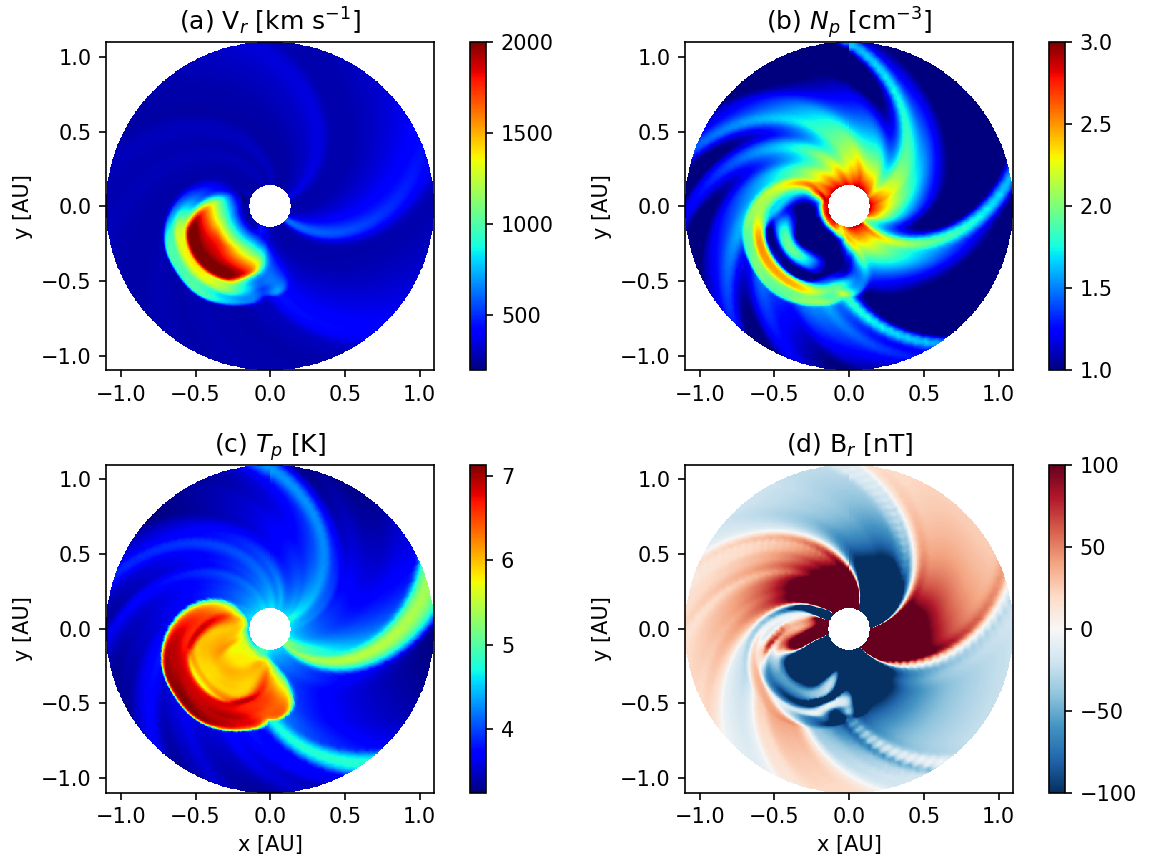}
    \caption{Equatorial slice of the CME simulation 20 hours after launch. Panels show: (a) radial velocity $V_r$ (km s$^{-1}$), (b) log of proton density $N_p$ (cm$^{-3}$), (c) log of proton temperature $T_p$ (K), and (d) radial magnetic field $B_r$ (nT).}
    \label{fig:July_2012_solar_storm}
\end{figure}

Figure~\ref{fig:July_2012_solar_storm} shows the simulation results 20 hours after the CME launch in the equatorial plane. Panel a) displays the radial velocity $V_r$ (km s$^{-1}$), revealing a strong forward shock that develops structure as it propagates through the CR2126 solar wind background. Panel b) shows the logarithm of the proton number density, $N_p$, (cm$^{-3}$), where a localized compression front is evident, potentially indicating the formation of a reverse shock. Panel c) presents the logarithm of proton temperature, $T_p$, (K), with a pronounced gradient consistent with shock heating. Panel d) shows the radial magnetic field, $B_r$, (nT), where the field structure inside the CME is distorted due to interaction with the ambient wind.

\subsection{Oscillatory Modes driven by CME}

\begin{figure*}
    \centering
    \includegraphics[width=\linewidth]{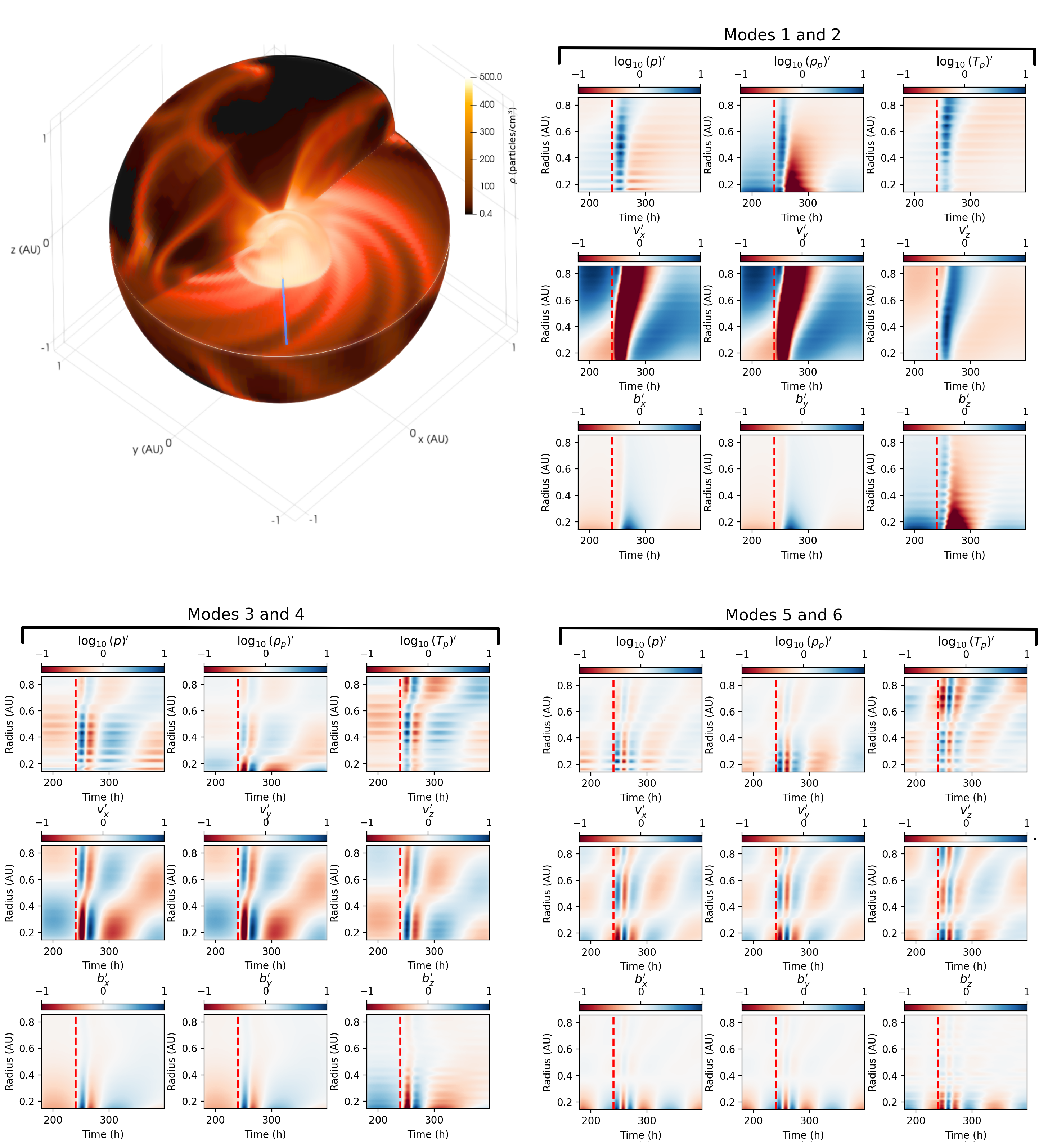}
    \caption{Top left panel: Three-dimensional view of the simulation domain, twelve hours after CME onset. The panel shows the (a) proton number density, volume-rendered and saturated at 500 particles/cm$^3$, highlighting the propagating CME structure. The blue line indicates the direction of CME propagation. Top right and bottom panels: Time–distance diagrams of the perturbation reconstructions for the variables $\log_{10}(p)$, $\log_{10}(N_p)$, $\log_{10}(T_p)$, $v_x$, $v_y$, and $v_z$, using modes 1–2, 3–4, and 5–6. The symbol $\prime$ denotes perturbed variables, and the time–distance profiles were computed along the blue line shown in the top left panel. The red dashed line marks the onset of the CME in the simulation  
    }
    \label{fig:CME-POD1-2}
\end{figure*}

The CME is launched 240 hours after the simulation starts, following a $\sim$230-hour relaxation period for the solar wind. Thus, the CME is injected approximately 10 hours after the solar wind reaches a steady state. The top left panel of Figure~\ref{fig:CME-POD1-2} shows a 3D snapshot at $t = 252$ hours, with volume-rendered number density (a) (saturated at 500 cm$^{-3}$) highlighting the CME structure. The blue line denotes the propagation direction of the CME.

To assess CME-driven dynamics, we apply vectorial SPOD to MHD variables as described in Eq. \ref{eq:vec-SPOD}, ensuring that intervariable correlations are preserved by ranking modes according to kinetic energy content. 
The analysis spans from $t = 180$ to 398 hours. The other panels in Figure~\ref{fig:CME-POD1-2} show time–distance diagrams along the blue line reconstructed using pairs of SPOD modes to represent perturbation propagation. The mode pairs were selected on the basis of the similarity of their spatial structures and the correlation of their temporal coefficients.  
For each pair of modes, the top row shows the perturbations of $\log_{10}(p)$, $\log_{10}(N_{p})$, and $\log_{10}(T_{p})$; the middle row displays the perturbations of the velocity components $v_x$, $v_y$, and $v_z$; and the bottom row presents the perturbations of the magnetic field components. The magnitudes of all physical variables were min-max normalized. From each reconstructed 2D time–distance diagram, we normalized the data and identified the wavefront as the first prominent peak in the radial profile at each time step using a peak detection algorithm. By tracking the position of this wavefront over time, we obtained a series of radius–time points.A linear fit to the wavefront trajectory was then applied to extract the average phase speed. For modes 1 and 2, a strong fast perturbation emerges in all variables, propagating at $>1000$ km s$^{-1}$ and persisting for $\sim$60 hours before dissipating. This is likely the signature of the CME initiating and propagating through the domain. The bottom panels show reconstructions of modes 3–4 and 5–6. Unlike the leading modes, the higher-order modes reveal two fast magnetoacoustic waves with initial speeds near 1000 km s$^{-1}$. Modes 3–4 correspond to a lower frequency wave with a period of approximately 213 hours, while modes 5-6 have a higher frequency with a period of approximately 128 hours. Both waves decelerate significantly after $\sim$24 hours, dropping below a speed of 500 km s$^{-1}$, suggesting
that the speed of the waves attenuates with the radial distance due to the spreading of energy over a larger and larger area as the wave propagates outwards.

\begin{figure}
    \centering
    \includegraphics[width=\columnwidth]{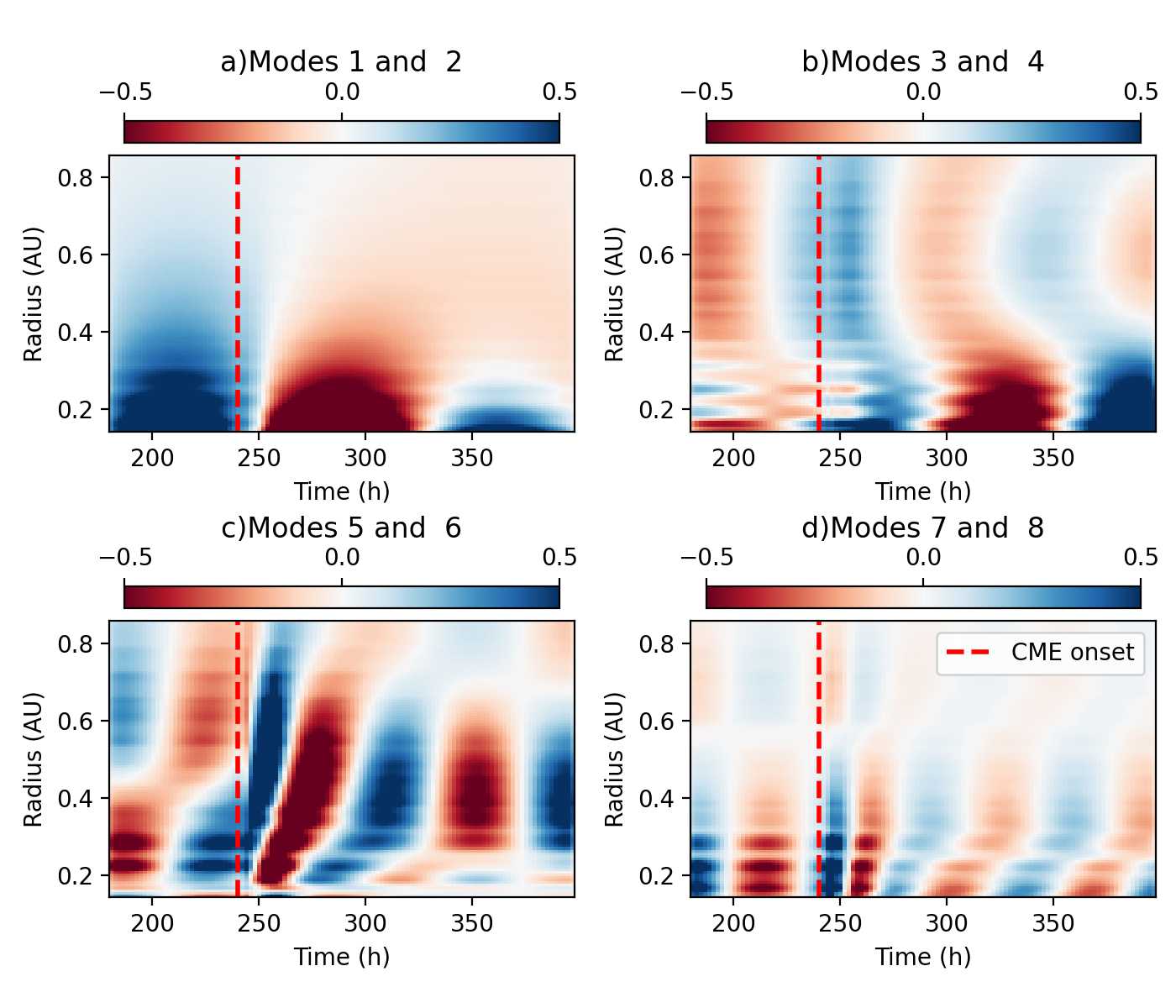}
    \caption{Time distance diagram along the blue line depicted in Figure \ref{fig:CME-POD1-2} of reconstructed perturbations using modes a) 1 and 2, b) 3 and 4, c) 5 and 6, and d) 7 and 8 of synthesized white light from simulations.} 
    \label{fig:CME-synthetic-time-coefficients}
\end{figure}

Similar wave signatures emerge when SPOD is applied to time series of synthetic white light, as indicated by perturbation reconstructions using modes 5–6 and 7–8 (Figure~\ref{fig:CME-synthetic-time-coefficients}), reinforcing the notion that fast magnetoacoustic waves observed in simulations should also be detectable in coronagraph data. The time distance diagrams in Figure~\ref{fig:CME-synthetic-time-coefficients} were computed along the blue line indicated in Fig~\ref{fig:CME-POD1-2}

\subsection{Wave detection in LASCO C3}

\begin{figure*}
    \centering
    \includegraphics[width=\linewidth]{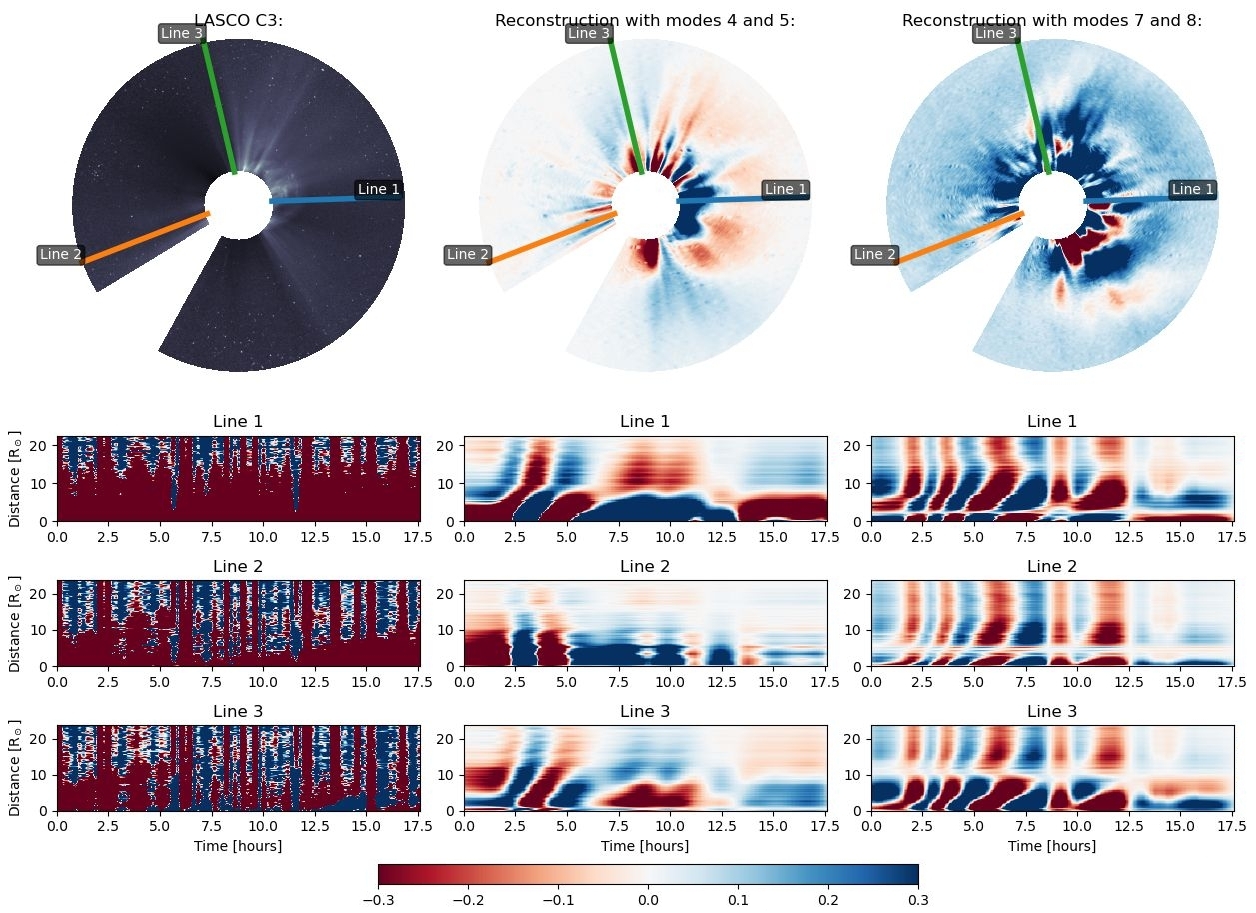}
    \caption{Analysis of LASCO C3 white image perturbations using selected modes of SPOD. The top panel shows the LASCO C3 image moments after the CME onset with identified lines of interest (1–3). The bottom panels display the corresponding time-distance diagrams for each line, where the vertical axis represents the distance from the Sun (in solar radii, R$_{\odot}$), and the horizontal axis indicates the time in hours. The CME onset occurs at $t=2$ hours. Red and blue hues denote positive and negative fluctuations in the SPOD intensity modes.}
    \label{fig:lasco-spod}
\end{figure*}

We applied SPOD to LASCO C3 white-light intensity perturbations to examine the CME evolution in observational data. Figure~\ref{fig:lasco-spod} displays a LASCO C3 image taken shortly after the CME onset (top left), along with the reconstructed perturbations using modes 4 and 5 (top middle) and modes 7 and 8 (top right). Three radial lines (labeled 1–3) are overlaid to guide the extraction of intensity profiles. The bottom panels show time–distance diagrams along each line, with distance in solar radii (R$_\odot$) and time in hours. The CME begins at $t = 2$ hours. The red and blue regions indicate SPOD-reconstructed positive and negative intensity perturbations, respectively, tracing CME-driven structures.

SPOD analysis of LASCO C3 perturbations reveals distinct propagation features by pairs of SPOD modes. For example, modes 4 and 5 show wavefronts travelling at 1200 km s$^{-1}$ and quickly fading away, while modes 7 and 8 display wavefronts decelerating from approximately 1200 to around 700 km s$^{-1}$ before vanishing. 
To quantify the dominant periodicity, we applied a Fast Fourier Transform (FFT) to the reconstructed intensity fluctuations along selected radial paths in the time–distance plots. The perturbations reconstructed by modes 4 and 5 have a dominant frequency of 0.082 mHz (period$\sim$ 3.39 hours) and persist for only one period. Modes 7 and 8 exhibit a higher frequency of 0.113 mHz (period$\sim$2.46 hours) and last for more than five hours. These findings suggest that lower-frequency perturbations exhibit more rapid attenuation than higher-frequency perturbations under the observed conditions.

The LASCO images analyzed with SPOD are based on Thomson-scattered visible light, which is directly proportional to the line-of-sight integrated electron density. Therefore, any perturbation identified in our time–distance or SPOD analysis must reflect changes in plasma density. As such, the observed propagating features are inherently compressive. Non-compressive modes, such as purely Alfv\'enic waves, do not produce detectable signatures in white-light coronagraph data, since they do not induce density variations measurable via Thomson scattering.

\begin{figure*}
    \centering
    \includegraphics[width=\linewidth]{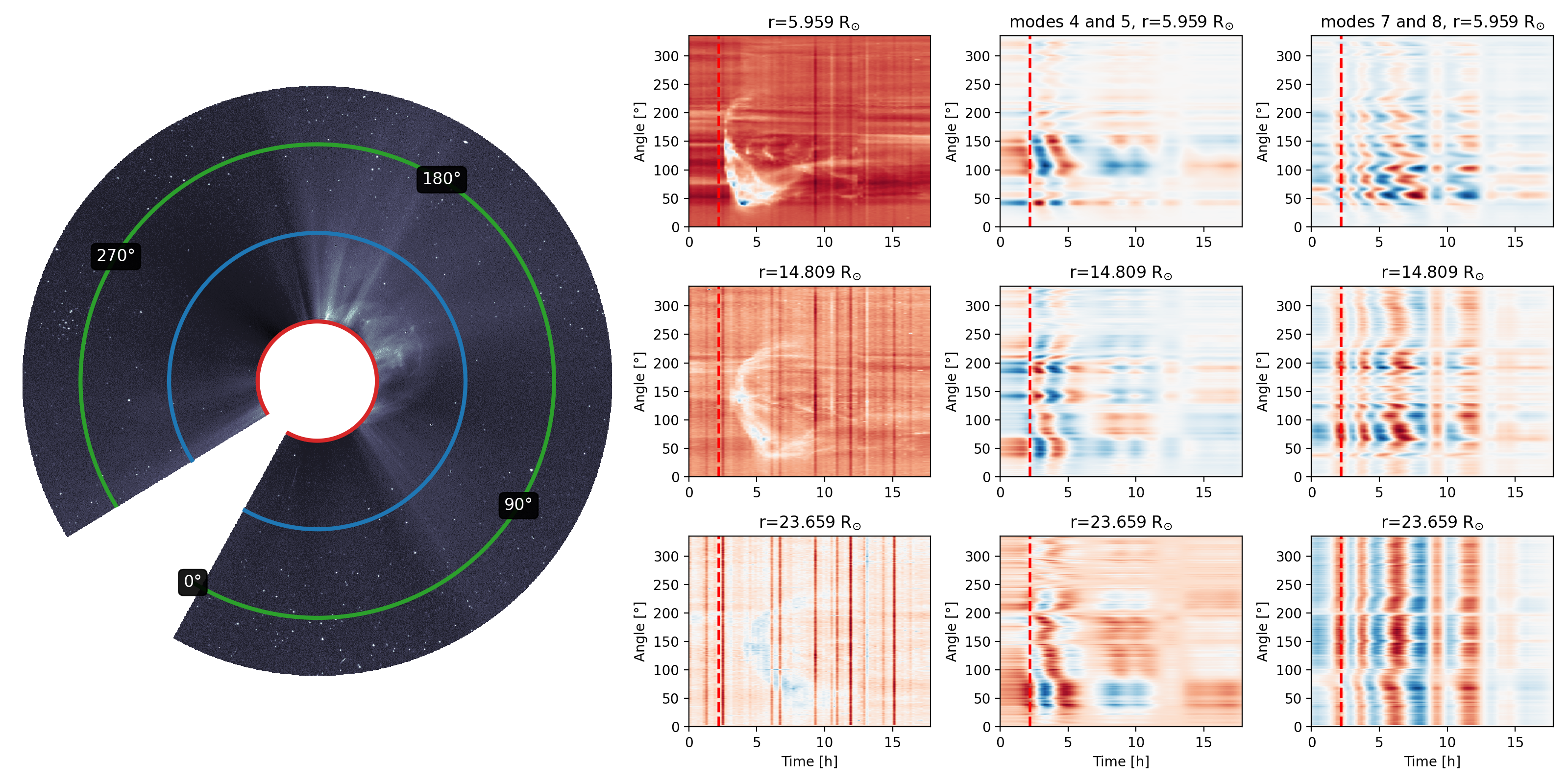}
    \caption{Analysis of LASCO C3 white image perturbations using selected modes of SPOD. The top panel shows a polar representation of the corona with concentric circles. The right panels display the corresponding time-distance diagrams for each circle, where the vertical axis represents the angle along the curve, and the horizontal axis indicates the time in hours. The CME onset occurs at $t=2$ hours. Red and blue hues denote positive and negative fluctuations in the SPOD intensity modes.}
    \label{fig:CME-LASCO}
\end{figure*}

Figure \ref{fig:CME-LASCO} displays the circular time-distance diagrams showing perturbations emerging at CME onset. The first panel shows the LASCO C3 image overplotted by the analysed circular radius: $r=5.95 R_\odot$ in red, $r=14.8R_\odot$ in blue, and $r=23.66R_\odot$ in green. Angles are also indicated. The second column shows the circular time-distance diagrams along those radii of the original LASCO C3 data for the analyzed time interval. The reconstructed perturbations obtained via POD and visualized using circular time–distance diagrams are shown in the third and fourth columns. Modes 4 and 5 (third column in Figure~\ref{fig:CME-LASCO} ) reconstruct a perturbation that exhibits a sharp increase in amplitude at different radii coinciding with the onset of the CME, consistent with impulsive and coherent energy release and the formation of a compressive front similar to a shock in the corona. Although this feature resembles a fast-mode shock, we note that in situ data from STEREO-A did not reveal a classical shock signature for this event \citep{Russell_2013}. This suggests that the observed coronal front may represent a transient, possibly subfast mode, or mediated shock that dissipated or did not persist to 1 AU. The subsequent decaying oscillations suggest compressional disturbances dissipating via geometrical spreading and energy losses in the expanding corona. The timing of the peak supports its association with a pressure discontinuity triggered by CME initiation.
This behavior supports the presence of a CME-driven compressive wavefront, consistent with a coronal shock that perturbs the corona and inner heliosphere. SPOD modes 7 and 8 (fourth column) reveal a second global circularly fast propagating compressive wave triggered across all radii at the onset of the CME, with oscillations lasting up to $\sim$12 hours. The coexistence of a rapidly propagating compressive front and a long-lived global mode suggests a dual-wave coronal response to the CME, involving both localized shock-like disturbances and global fast-mode wave activity.

\section{Discussions and conclusion}

We presented the first combined observational and numerical evidence for a dual-wave response to a CME in the outer corona and inner heliosphere. We identified a rapidly propagating compressive front consistent with a fast-mode shock and a persistent global coronal oscillation. This dual behavior is evident in the SPOD-decomposed LASCO data and an MHD simulation, with each mode exhibiting distinct spatial and temporal evolution patterns. The compressive front extends approximately 60$^\circ$ to 90$^\circ$ around the CME nose, while the global mode spans nearly the full 360$^\circ$ angular extent of the C3 field of view.

This result provides new support for the idea that CMEs distribute energy into multiple MHD wave components. The global mode is consistent with previous interpretations of large-scale coronal oscillations and EUV waves \citep[e.g.,][]{Liu_2018, Zheng_2019}, while the compressive front aligns with piston- and bow-shock models \citep{Liu_2017}. The ability of the SPOD method to isolate these components confirms its value in disentangling overlapping wave structures in coronagraph data and demonstrates its consistency with controlled numerical experiments.

The CME-driven compressive front, observed for $\sim$2.5 hours in LASCO data, shows propagation speeds exceeding 1000~km~s$^{-1}$, similar to fast-mode EUV waves and white-light shock signatures \citep{Gopalswamy2008, Ontiveros2009}. In the simulation, this front persists for $\sim$60 hours and reaches 1~AU, though its morphology varies with distance. However, in situ observations from STEREO-A \citep{Russell_2013} show no clear fast-mode shock signature, suggesting that the shock-like disturbance we observe in the outer corona may have been suppressed or mediated by upstream conditions; consistent with scenarios of shock weakening or dissipation in the heliosphere \citep[e.g.,][]{Baker2013, Riley_et_al_2016}.

Although our analysis is limited to white-light coronagraph data, the results provide evidence for global, CME-driven coronal oscillations analogous to, and likely the high-altitude extension of, EUV waves described in previous studies (e.g., \cite{Hu_2024}). This suggests that CMEs can excite surface-propagating EUV waves and large-scale compressive modes extending far beyond the traditional EUV domain, as seen in the nearly spherical shocks reported by \cite{Liu_2017} for the same event. Thus, such large-scale oscillations may be a common but underexplored feature of major solar eruptions, with important implications for the structure of the heliosphere and space weather prediction. Moreover, the SPOD method provides a new diagnostic capability for tracing how CME energy couples into fast-mode wave activity across large spatial scales.

\section*{Acknowledgments}  
S.S.A.S., V.F., G.V. and I.B. are grateful to The Royal Society, International Exchanges Scheme, collaboration with Greece (IES/R1/221095) and India (IES/R1/211123). S.S.A.S., V.F., and G.V. are grateful for the Science and Technology Facilities Council (STFC) grants ST/V000977/1, ST/Y001532/1.  L.A.C.A.S.,  acknowledge STFC for support from grant No. ST/X001008/1. L.F.G.B. was supported by Brazilian agency CAPES under grant (88887.006144/2024-00 - 2200101-8 UFC). E.L.R. acknowledges support by Brazilian agency CNPq under grant 306920/2020-4. S.S.A.S., V.F., and E.L.R. would like to thank the International Space Science Institute (ISSI) in Bern, Switzerland, for the hospitality provided to the team on "Opening New Avenues in Identifying Coherent Structures and Transport Barriers in the Magnetised Solar Plasma." V.F. also acknowledges ISSI's support for the teams on "The Nature and Physics of Vortex Flows in Solar Plasmas," and “Tracking Plasma Flows in the Sun’s Photosphere and Chromosphere: A Review \& Community Guide.”. 
J.J.G.A. acknowledges the support of the \textit{Secretaría de Ciencia, Humanidades, Tecnología e Innovación} (SECIHTI) under grant 319216, \textit{Modalidad: Paradigmas y Controversias de la Ciencia 2022}, and UNAM-PAPIIT grant IA100725. P.R. and M.B.N. acknowledge support from NSF's PREEVENTS program (grant no.\ ICER-1854790), NASA's LWS Strategic Capabilities (80NSSC22K0893), and the PSP WISPR contract NNG11EK11I to NRL (under subcontract N00173-19-C-2003 to PSI). 

\bibliographystyle{aasjournal}
\bibliography{CME-Driven_Shock_and_Global_Oscillation}

\end{document}